\documentclass[pra,onecolumn,superscriptaddress]{revtex4-1}

\usepackage{amssymb}
\usepackage{graphicx}
\usepackage{bbm,amsmath,amssymb}
\usepackage{array}

\usepackage{amsthm}
\usepackage{graphicx}
\usepackage[usenames,dvipsnames]{color}
\usepackage{epsfig,color}
\usepackage{amsmath,amssymb}
\usepackage[T1]{fontenc}
\usepackage{verbatim}
\usepackage{float}
\usepackage{units}
\usepackage{amsmath}
\usepackage{graphicx}
\usepackage[colorlinks = true,
            linkcolor = NavyBlue,
            urlcolor  = NavyBlue,
            citecolor = NavyBlue,
            anchorcolor = blue]{hyperref}
\usepackage{epstopdf}

\makeatletter

 \@ifundefined{textcolor}{}
 {
   \definecolor{BLACK}{gray}{0}
   \definecolor{WHITE}{gray}{1}
   \definecolor{RED}{rgb}{1,0,0}
   \definecolor{GREEN}{rgb}{0,1,0}
   \definecolor{BLUE}{rgb}{0,0,1}
   \definecolor{CYAN}{cmyk}{1,0,0,0}
   \definecolor{MAGENTA}{cmyk}{0,1,0,0}
   \definecolor{YELLOW}{cmyk}{0,0,1,0}
 }

\makeatother

\begin{document}

\title{Quantum Simulation of Dissipative Processes without Reservoir Engineering}

\author{R. Di Candia}
\email{rob.dicandia@gmail.com}
\affiliation{Department of Physical Chemistry, University of the Basque Country UPV/EHU, Apartado   644, 48080 Bilbao, Spain}
\author{J. S. Pedernales}
\affiliation{Department of Physical Chemistry, University of the Basque Country UPV/EHU, Apartado   644, 48080 Bilbao, Spain}
\author{A. del Campo}
\affiliation{Department of Physics, University of Massachusetts, Boston, MA 02125, USA}
\affiliation{Theoretical Division,  Los Alamos National Laboratory, Los Alamos, NM 87545, USA}
\affiliation{Center for Nonlinear Studies,  Los Alamos National Laboratory, Los Alamos, NM 87545, USA}
\author{E. Solano}
\affiliation{Department of Physical Chemistry, University of the Basque Country UPV/EHU, Apartado   644, 48080 Bilbao, Spain}
\affiliation{IKERBASQUE, Basque Foundation for Science, Maria Diaz de Haro 3, 48013 Bilbao, Spain}
\author{J. Casanova}
\affiliation{Department of Physical Chemistry, University of the Basque Country UPV/EHU, Apartado   644, 48080 Bilbao, Spain}
\affiliation{Institut f\"{u}r Theoretische Physik, Albert-Einstein-Allee 11, Universit\"{a}t Ulm, D-89069 Ulm, Germany}

\begin{abstract}
We present a quantum algorithm to simulate general finite dimensional Lindblad master equations without the requirement  of  engineering the system-environment interactions. The proposed method is able to simulate both Markovian and non-Markovian quantum dynamics. It consists in the quantum computation of the dissipative corrections to the unitary evolution of the system of interest, via the reconstruction of the response functions associated with the Lindblad operators.  Our approach is equally applicable to dynamics generated by effectively non-Hermitian Hamiltonians. We confirm the quality of our method providing specific error bounds that quantify  its  accuracy.
\end{abstract}

\pacs{03.67.Ac, 03.65.Yz}

\maketitle

While every physical system is indeed coupled to an environment~\cite{BP02, RH11}, modern quantum technologies have succeeded in isolating systems to an exquisite degree in a variety of platforms~\cite{LeibfriedEtAl, Devoret13, Bloch05, Obrien09}. In this sense, the last decade has witnessed great advances in testing and  controlling the quantum features of these  systems, spurring 
the quest for the development of quantum simulators \cite{Feynman82, Lloyd96, CZ12,Nori14}. These efforts are guided by the early proposal 
of using a highly tunable quantum device to mimic  the behavior of another quantum system of interest,  being  the latter complex enough to render its description by classical means intractable.
By now, a series of proof-of-principle experiments have  successfully demonstrated the basic tenets of quantum simulations revealing quantum technologies as  trapped ions \cite{qsions}, ultracold quantum gases \cite{uca}, and superconducting circuits \cite{qsscc}    as  promising candidates to harbor  quantum simulations beyond the computational capabilities of classical devices.

It was soon recognised that this endeavour should not  be limited to simulating the dynamics of isolated complex  quantum systems, 
but should more generally aim at the emulation of  arbitrary physical processes, 
including the open quantum dynamics of  a system coupled to an environment.
Tailoring the complex nonequilibrium dynamics of an open system has the potential to uncover a plethora of technological and scientific applications.
A remarkable instance results from the understanding of the role played by quantum effects in the open dynamics of photosynthetic processes in biological systems \cite{HP13,Mostame12}, 
recently used in the design of artificial light-harvesting nanodevices \cite{Scully11,Dorfman13,Creatore13}. At a more fundamental level, an open-dynamics quantum simulator would be invaluable to shed new light on core issues of foundations of physics, ranging from the quantum-to-classical transition  and quantum measurement theory \cite{Zurek03} to the characterization of Markovian and non-Markovian systems \cite{NM1,NM2,NM3}.
Further motivation arises at the forefront of quantum technologies.  As  the available resources increase,  the verification with classical computers of quantum annealing devices \cite{Boixo13,Boixo13b}, possibly operating with a  hybrid quantum-classical performance, becomes a daunting task. The comparison between different experimental implementations of quantum simulators is required to establish a confidence level, as customary with other quantum technologies, e.g., in the use of atomic clocks for time-frequency standards. In addition, the knowledge and control of dissipative processes can be used as well as a resource for quantum state engineering~\cite{VWC09}. 

Facing  the high dimensionality of the Hilbert space of the composite system made of a quantum device embedded in an environment, 
recent developments have been focused on the reduced  dynamics of the system that emerges after  tracing out the environmental degrees of freedom.
The resulting nonunitary dynamics is governed by a dynamical map,  or equivalently,  by a master equation~\cite{BP02,RH11}.
In this respect, theoretical~\cite{LV01, Eisert, Nori11} and experimental~\cite{Barreiro11} efforts in the simulation of open quantum systems have exploited the combination of coherent quantum operations 
with controlled dissipation.
Notwithstanding, the experimental complexity required to simulate an arbitrary open quantum dynamics is recognised to substantially surpass that needed in the case of  closed systems,  
where a smaller number of generators suffices to design a general time-evolution. Thus, the quantum simulation of open systems remains a challenging task.

In this Letter, we propose a quantum algorithm to simulate finite dimensional Lindblad master equations, corresponding to Markovian or non-Markovian processes. Our protocol shows how to reconstruct, up to an arbitrary finite error, physical observables that evolve according to a dissipative dynamics, by evaluating multi-time correlation functions of its Lindblad operators. We show that the latter requires the implementation of the unitary part of the dynamics in a quantum simulator, without the necessity of physically engineering the system-environment interactions.  Moreover, we demonstrate  how these multi-time correlation functions can be computed  with a reduced number of measurements. We further show that our method can be applied as well to the simulation of  processes associated with non-Hermitian Hamiltonians. Finally, we provide specific error bounds to estimate the accuracy of our approach. 

Consider a quantum system coupled to an environment whose dynamics is described by  the von Neumann equation  $ i\frac{d{\bar{\rho}}}{dt} = [\bar H, \bar{\rho}]$. Here, $\bar{\rho}$ is the system-environment density matrix, $\bar{H}= H_s + H_e + H _I$,   where $H_s$ and $H_e$  are the  system and environment Hamiltonians, while  $H_I$ corresponds to  their interaction. Assuming weak coupling and short time-correlations between the system and the environment, after tracing out the environmental degrees of freedom we obtain the Markovian master equation
\begin{equation}\label{master}
\frac{d\rho}{dt} =\mathcal{L}^t_{}\rho,
\end{equation}
being $\rho={\rm Tr}_e(\bar{\rho})$ and $\mathcal{L}^t$  the time-dependent superoperator governing the dissipative dynamics~\cite{BP02, RH11}.  Notice that there are different ways to recover Eq.~\eqref{master} \cite{Alicki06}. Nevertheless, Eq.~\eqref{master} is our starting point, and in the following we show how to simulate this equation regardless of its derivation. Indeed, our algorithm does not need control any of the approximations done to achieve this equation. We can decompose  $\mathcal{L}^t$ into $\mathcal{L}^t=\mathcal{L}_{H}^t+\mathcal{L}_D^t$. Here, $\mathcal{L}_H^t$ corresponds to a unitary part, i.e. $\mathcal{L}_{H}^t\rho\equiv-i[H(t),\rho]$, where $H(t)$ is defined by $H_s$ plus a term due to the lamb-shift effect and it may depend on time. Instead, $\mathcal{L}_D^t$ is the dissipative contribution and it follows the Lindblad form~\cite{Lindblad} $\mathcal{L}_D^t\rho\equiv\sum_{i=1}^N\gamma_i(t)\left(L_i\rho L_i^\dag-\frac{1}{2}\{L_i^{\dag}L_i, \rho\}\right)$, where $L_i$ are the Lindblad operators modelling the effective interaction of the system with the bath that may depend on time, while $\gamma_i(t)$ are nonnegative parameters.  Notice that, although the standard derivation of Eq.~\eqref{master} requires the Markov approximation, a non-Markovian equation can have the same form. Indeed, it is known that if $\gamma_i(t)<0$ for some $t$ and $\int_0^tdt'\;\gamma_i(t')>0$ for all $t$, then Eq.~\eqref{master} corresponds to a completely positive non-Markovian channel~\cite{Rivas14}. Our approach can deal also with non-Markovian processes of this kind, keeping the same efficiency as the Markovian case. While we will consider the general case $\gamma_i=\gamma_i(t)$, whose sign distinguishes the Markovian processes by the non-Markovian ones, for the sake of simplicity we will consider the case $H\neq H(t)$ and $L_i\neq L_i(t)$ (in the following, we will denote $\mathcal{L}_H^t$ simply as $\mathcal{L}_H$). However, the inclusion in our formalism of  time-dependent Hamiltonians and Lindblad operators is straightforward.

One can integrate Eq.~\eqref{master} obtaining a Volterra equation~\cite{Bellmann}  
\begin{align}\label{Volterra}
\rho(t)=e^{t\mathcal{L}_{H}}\rho(0)+\int_0^tds\;e^{(t-s)\mathcal{L}_{H}}\mathcal{L}_D^s\ \rho(s),
\end{align}
where $e^{t\mathcal{L}_{H}}\equiv\sum_{k=0}^\infty t^k\mathcal{L}_{H}^k/k!$.  The first term at the right-hand-side of Eq.~(\ref{Volterra}) corresponds to the  unitary evolution of $\rho(0)$ while the second term gives rise to the dissipative correction. Our goal is to find a perturbative expansion of Eq.~\eqref{Volterra} in the $\mathcal{L}_D^t$ term, and to provide with a  protocol to measure the resulting expression in a unitary way. In order to do so, we consider the iterated solution of Eq.~(\ref{Volterra}) obtaining 
\begin{align}
\rho(t)\equiv  \sum_{i=0}^\infty \rho_i(t).\label{series}
\end{align}
Here, $\rho_0(t)=e^{t\mathcal{L}_{H}}\rho(0)$, while, for $i\geq1$, $\rho_i(t)$ has the following general structure: $\rho_i(t)= \Pi_{j=1}^i  \Phi_j  \ e^{s_i \mathcal{L}_{H}} \rho(0)$,
$\Phi_j$ being a superoperator acting on an arbitrary matrix $\xi$ as  $\Phi_j \xi = \int_0^{s_{j-1}}ds_{j} \ e^{(s_{j-1} - s_{j}) \mathcal{L}_{H}}\mathcal{L}_D^{s_{j}}\, \xi$,
where $s_0\equiv t$. For instance,  $\rho_2(t)$ can be written as 
\begin{equation}
\rho_2(t) = \Pi_{j=1}^2 \Phi_j   e^{s_2 \mathcal{L}_{H}} \rho(0) = \Phi_1\Phi_2\ e^{s_2 \mathcal{L}_{H}} \rho(0) = \int_0^{t}ds_1e^{(t-s_1)\mathcal{L}_{H}}\mathcal{L}_D^{s_1}\int_{0}^{s_1}ds_2 e^{(s_1-s_2)\mathcal{L}_{H}}\mathcal{L}_D^{s_2}  e^{s_2\mathcal{L}_{H}}\rho(0).\nonumber
\end{equation}
In this way, Eq.~(\ref{series}) provides us with a general and useful expression of the solution of Eq.~\eqref{master}. Let us consider the truncated series in Eq.~(\ref{series}), that is $\tilde\rho_{n}(t) = e^{t\mathcal{L}_{H}}\rho(0) +  \sum_{i=1}^{n} \rho_i(t)$, where $n$  corresponds to the order of the approximation. We will prove that an expectation value $\langle O \rangle_{\rho(t)}\equiv\text{Tr}\,[O\rho(t)]$ corresponding to a dissipative dynamics can be well approximated as 
\begin{equation}\label{apprO}
\langle O\rangle_{\rho(t)}\approx {\rm Tr}[O e^{t\mathcal{L}_{H} } \rho(0)]  + \sum_{i=1}^{n} {\rm Tr}[O \rho_i(t)].
\end{equation}
In the following, we will supply with a quantum algorithm based on single-shot random measurements to compute each of the terms appearing in Eq.~(\ref{apprO}), and we will derive specific upper-bounds quantifying the accuracy of  our method. Notice that the first term at the right-hand-side of  Eq.~(\ref{apprO}), i.e. ${\rm Tr}[O e^{t\mathcal{L}_{H} } \rho(0)]$,  corresponds to the expectation value of the operator $O$ evolving under a unitary dynamics, thus it can be measured directly in a unitary quantum simulator where the dynamic associated with the Hamiltonian $H$ is implementable. However, the successive terms of the considered series, i.e. ${\rm Tr}[O \rho_i(t)]$ with $i\geq1$, require a specific development because they involve  multi-time correlation functions of the Lindblad operators and the operator $O$. 

Let us consider the first order term of the series in Eq.~\eqref{apprO}
\begin{equation}\label{first}
\langle O\rangle_{\rho_1(t)}=\int_0^tds_1\; \text{Tr}\,[Oe^{(t-s_1)\mathcal{L}_{H}}\mathcal{L}_D^{s_1} \rho_0(s_1)]=\sum_{i=1}^N\int_0^tds_1\;\gamma_i(s_1)\bigg[\langle L_i^\dag (s_1)O(t)L_i (s_1)\rangle-\frac{1}{2}\langle \left\{O(t), L_i^\dag L_i(s_1)\right\}\rangle\bigg],
\end{equation}
where $\xi(s)\equiv e^{iH s}\xi e^{-iH s}$ for a general operator $\xi$ and time $s$, and all the expectation values are computed in the state $\rho(0)$. Note that the average values  appearing in the second and third lines of  Eq.~\eqref{first} correspond to time correlation functions of the operators $O$, $L_i^{}$, $L^{\dag}_i$, and $L_i^\dag L_i$. In the following, we consider a basis $\{Q_{j}\}_{j=1}^{d^2}$, where $d$ is the system dimension and $Q_{j}$  are Pauli-kind operators, i.e. both unitary and Hermitian (see supplemental material~\cite{suppl} for more details). The operators $L_i$ and $O$ can be decomposed as $L_i=\sum_{k=1}^{M_i} q^i_{k}Q^i_{k}$ and $O=\sum_{k=1}^{M_O} q^O_{k}Q^O_{k}$, with $q^{i,O}_{k}\in\mathbb{C}$, $Q^{i,O}_{k}\in \{ Q_j\}_{j=1}^{d^2}$, and $M_i,M_O\leq d^2$. We obtain then
\begin{equation}\label{decomp}
\langle L_i^\dag(s_1) O(t)L_i(s_1) \rangle=\sum_{l=1}^{M_{O}}\sum_{k,k'=1}^{M_i}q_l^{O}q_{k}^{i\,*}q_{{k'}}^i\langle Q^i_{k}(s_1) Q_l^O(t)Q_{{k'}}^i(s_1)\rangle,
\end{equation}
that is a sum of correlations of unitary operators. The same argument applies to the terms including $L_i^\dag L_i$ in  Eq.~\eqref{first}. Accordingly, we have seen that the problem of estimating the first-order correction is moved to the  measurement of some specific multi-time correlation functions involving the $Q_k^{i,O}$ operators. The argument can be easily extended to higher-order corrections. Indeed, for the $n$-th order, we have to evaluate the quantity
\begin{equation}\label{nquant}
\langle O\rangle_{\rho_n(t)}=\int dV_n\;\text{Tr}[Oe^{(t-s_1)\mathcal{L}_{H}}\mathcal{L}_D^{s_1} \dots  \mathcal{L}_D^{s_n}e^{s_n\mathcal{L}_{H}}\rho(0)]\equiv\sum_{i_1,\dots,i_n=1}^N\int dV_n\;\langle A_{[i_1,\cdots,i_n]}(\vec{s})\rangle.
\end{equation}
Here,
\begin{align}
A_{[i_1,\dots,i_n]}(\vec{s})&\equiv  e^{s_n\mathcal{L}_{H}^\dag}\mathcal{L}_D^{s_n,i_n\dag}\dots \mathcal{L}_D^{s_2,i_2\dag}e^{(s_1-s_2)\mathcal{L}^\dag_{H}} \mathcal{L}_D^{s_1,i_1\dag}e^{(t-s_1)\mathcal{L}^\dag_{H}}O,\nonumber
\end{align}
where $\mathcal{L}_{D}^{s,i} \xi\equiv \gamma_{i}^{}(s)\left(L_{i}\xi L_{i}^{\dag}-\frac{1}{2}\{ L^{\dag}_{i} L_{i}, \xi\}\right)$,  $\vec{s} = (s_1, \dots, s_n)$, $\int dV_n = \int_0^t \dots \int_0^{s_{n-1}} ds_1\dots ds_{n}$, and $\mathcal{L}^\dag\xi\equiv(\mathcal{L}\xi)^\dag$ for a general superoperator $\mathcal{L}$. As  in Eq.~(\ref{first}), the above expression contains multi-time correlation functions of the Lindblad operators $L_{i_1},\dots, L_{i_n}$ and the observable $O$, that have to be evaluated in order to compute  each  contribution  in Eq.~(\ref{apprO}). 

Our next step is to provide a method to evaluate general terms as the one appearing in Eq.~\eqref{nquant}. The standard approach to estimate this kind of quantities corresponds to measuring the expected value $\langle A_{[i_1,\cdots,i_n]}(\vec{s})\rangle$ at different random times $\vec{s}$ in the integration domain, and then calculating the average. Nevertheless, this strategy involves a huge number of measurements, as we need to estimate an expectation value at each chosen time. Our technique, instead, is based on single-shot random measurements and, as we will see below, it leads to an accurate estimate of  Eq.~(\ref{nquant}). More specifically, we will prove that 
\begin{equation}\label{casilla}
\sum_{i_1,\dots,i_n=1}^N\int dV_n\;\langle A_{[i_1,\cdots,i_n]}(\vec{s})\rangle\approx \frac{N^n|V_n|}{|\Omega_n|} \sum_{\Omega_n}\tilde A_{\vec{\omega}}(\vec{t}),
\end{equation}
where $\tilde A_{\vec{\omega}}(\vec{t})$ corresponds to a single-shot measurement of $A_{\vec{\omega}}(\vec{t})$, being  $[\vec{\omega},\vec{t}]\in\Omega_n\subset\{[\vec{\omega},\vec{t}]\;|\; \vec{\omega}=[i_1,\dots, i_n], i_k\in[1,N], \vec{t}\in V_n \}$, $|\Omega_n|$ is the size of $\Omega_n$, and $[\vec{\omega},\vec{t}]$ are sampled uniformly and independently. As already pointed out, the integrand in Eq.~\eqref{nquant} involves multi-time correlation functions. In this respect, we note that  a quantum algorithm for their efficient reconstruction has recently been  proposed~{\cite{Pedernales}}. Indeed, the authors in Ref.~{\cite{Pedernales}} show how, by adding only one ancillary qubit to the simulated system,  general time-correlation functions are accessible by implementing only unitary evolutions of the kind $e^{t\mathcal{L}_{H}}$, together with entangling operations between the ancillary qubit and the system. It is noteworthy to mention that these operations have already experimentally demonstrated in quantum systems as trapped ions~\cite{Lanyon11} or quantum optics~\cite{Obrien09}, and have been recently proposed for cQED architectures~\cite{Mezzacapo14}. Moreover, the same quantum algorithm allows us to measure single-shots of the real and imaginary part of these quantities providing, therefore, a way to compute the term at the right-hand-side of Eq.~(\ref{casilla}). Notice that the evaluation of each term $\langle A_{[i_1,\cdots,i_n]}(\vec{s})\rangle$ in Eq.~\eqref{nquant}, requires a number of measurements that depends on the observable decomposition, see Eq.~\eqref{decomp}. After specifying it, we measure the real and the imaginary part of the corresponding correlation function. Finally, in the supplemental material~\cite{suppl} we prove that
\begin{align}\label{numer1}
\left|\sum_{i_1,\dots,i_n=1}^N\int dV_n\;\langle A_{[i_1,\cdots,i_n]}(\vec{s})\rangle-\frac{(Nt)^n}{n!|\Omega_n|}\sum_{\Omega_n}\tilde A_{\vec{\omega}}(\vec{t})\right|\leq\delta_n
\end{align}
with probability higher than $1-e^{-\beta}$, provided that $|\Omega_n|>\frac{36M_O^2(2+\beta)}{\delta_n^2}\frac{(2\bar{\gamma} MN t)^{2n}}{n!^2}$, where $\bar{\gamma}\equiv\max_{i,s\in[0,t]}|\gamma_i(s)|$ and $M\equiv\max_i M_i$. Equation~\eqref{numer1} means that   that  the quantity in Eq.~\eqref{nquant} can be estimated with arbitrary precision by random single-shot measurements of $A_{[i_1,\cdots,i_n]}(\vec{s})$, allowing, hence, to dramatically reduce  the resources required by our quantum simulation algorithm. Notice that the required number of measurements to evaluate the order $n$ is bounded by $3^n|\Omega_n|$, and the total number of measurements needed to compute the correction to the expected value of an observable up the order $K$ is bounded by $\sum_{n=0}^K3^n|\Omega_n|$. In the following, we discuss at which order we need to truncate in order to have a certain error in the final result.
 
So far, we have proved that we can compute, up to an arbitrary order in $\mathcal{L}_D^t$, expectation values corresponding to dissipative dynamics with a unitary quantum simulation. It is noteworthy  that our method does not require to physically engineer  the system-environment interaction. Instead, one only needs to implement the system Hamiltonian $H$. In this way we are opening a new avenue for the quantum simulation of open quantum dynamics in situations where the complexity on the design of the dissipative terms excedes the capabilities of quantum platforms. This covers a wide range of physically relevant situations. One example corresponds to the case of fermionic theories where the encoding of the fermionic behavior in the  degrees of freedom of the quantum simulator gives rise to highly delocalized operators~\cite{JordanWigner, Casanova12}. In this case a reliable dissipative term should act on these non-local operators instead of on the individual qubits of the system. Our protocol solves this problem because it avoids the necessity of implementing the Lindblad superoperator. Moreover, the scheme allows one to simulate at one time a class of master equations corresponding to the same Lindblad operators, but with different choices of $\gamma_i$, including the relevant case when only a part of the system is subjected to dissipation, i.e. $\gamma_i=0$ for some values of $i$.

We shall next quantify the quality of our method. In order to do so, we will find an error bound certifying how the truncated series in Eq.~\eqref{series} is close to the solution of Eq.~\eqref{master}. This error bound will depend on the system parameters, i.e. the time $t$ and the dissipative parameters $\gamma_i$. As figure of merit we choose the trace distance, defined by
\begin{equation}
D_1(\rho_1,\rho_2)\equiv \frac{\|\rho_1-\rho_2\|_1}{2},
\end{equation}
where $\|A\|_1\equiv\sum_i\sigma_i(A)$, being $\sigma_i(A)$ the  singular values of $A$~\cite{Watrous}. Our goal is to find a bound for $D_1(\rho(t),\tilde {\rho}_n(t))$, where $\tilde \rho_n(t)\equiv \sum_{i=0}^n\rho_i(t)$ is the series of Eq. \eqref{series} truncated at the $n$-th order. We note that the the following recursive relation holds 
\begin{equation}\label{generaldif}
{\tilde {\rho}}_n(t)=e^{t\mathcal{L}_H}\rho(0)+\int_0^tds\;e^{(t-s)\mathcal{L}_{H}}\mathcal{L}_D^s\tilde{\rho}_{n-1}(s).
\end{equation}
From Eq. \eqref{generaldif}, it follows that
\begin{equation}\label{induce}
D_1(\rho(t),\tilde{\rho}_n(t))=\frac{1}{2}\left\|\int_0^tds\;e^{(t-s)\mathcal{L}_{H}} \mathcal{L}_D^s(\rho(s)-\tilde{\rho}_{n-1}(s))\right\|_1\leq\int_0^tds\;\|\mathcal{L}_D^s\|_{1\rightarrow1}D_1(\rho(s),\tilde{\rho}_{n-1}(s)),
\end{equation}
where we have introduced the induced superoperator norm $\| \mathcal{A}\|_{1\rightarrow 1}\equiv\sup_\sigma\frac{\|\mathcal{A}\sigma\|_1}{\|\sigma\|_1}$~\cite{Watrous}. For $n=0$, i.e. for $\tilde \rho_n(t)\equiv \tilde{\rho}_0(t) = e^{t\mathcal{L}_{H}}\rho(0)$, we obtain the following bound
\begin{align}\label{D1zero}
D_1(\rho(t),\tilde{\rho}_0(t))&\leq \frac{1}{2}\int_0^tds\; \|\mathcal{L}_D^s\|_{1\rightarrow1}\|\rho(s)\|_1\leq \sum_{i=1}^N |\gamma_i(\epsilon_i)|\|L_i\|_\infty^2t,
\end{align}
where $0\leq \epsilon_{i}\leq t$~\cite{suppl}, and $\|A\|_\infty \equiv\sup_i\sigma_i(A)$. Notice that, in finite dimension, one can always renormalize $\gamma_i$ in order to have $\|L_i\|_{\infty}=1$,  i.e. if we transform $L_i\rightarrow L_i / \|L_i\|_\infty$, $\gamma_i\rightarrow \|L_i\|_\infty \gamma_i$, the master equation remains invariant. Using Eq.~\eqref{induce}-\eqref{D1zero}, one can shown by induction that for the general $n$-th order the following bound holds 
\begin{equation}\label{bounds}
D_1(\rho(t),\tilde{\rho}_n(t))\leq\prod_{k=0}^n\bigg[2\sum_{i_k=1}^N |\gamma_{i_k}(\epsilon_{i_k})|\bigg]\frac{t^{n+1}}{2(n+1)!}\leq\frac{(2\bar \gamma N t)^{n+1}}{2(n+1)!},
\end{equation}
where $0\leq \epsilon_{i_k}\leq t$ and we have set $\| L_i\|_\infty=1$. From Eq.~\eqref{bounds}, it is clear that the series converges uniformly to the solution of Eq.~\eqref{master} for every finite value of $t$ and choices of $\gamma_i$. As a result, the number of measurements needed to simulate a certain dynamics at time $t$ up to an error $\varepsilon<1$ is $O\left(\left(\bar{t}+\log\frac{1}{\varepsilon}\right)^2\frac{e^{12M\bar{t}}}{\varepsilon^2}\right)$, where $\bar{t}=\bar \gamma N t$~\cite{suppl}. Here, a discussion on the efficiency of the method is needed. From the previous formula, we can say that our method performs well when $M$ is low, i.e. in that case where each Lindblad operators can be decomposed in few Pauli-kind operators. Moreover, as our approach is perturbative in the dissipative parameters $\gamma_i$, it is reasonable that the method is more efficient when $|\gamma_i|$ are small. Notice that analytical perturbative techniques are not available in this case, because the solution of the unperturbed part is assumed to be not known. Lastly, it is evident that the algorithm is efficient for a certain choices of time, and the relevance of the simulation depends on the particular cases. For instance, a typical interesting situation is a strongly coupled Markovian system. Let us assume  with site-independent couple parameter $g$ and dissipative parameter $\gamma$. We have that $e^{12M\bar t}\leq 1+12eM\bar t$ if $t\leq \frac{1}{12M\gamma N}\equiv t_c$. In this period, the system oscillates typically $C\equiv gt_c=\frac{g/\gamma}{12MN}$ times, so the simulation can be considered efficient for $N\sim g/\gamma C$, which, in the strong coupling regime, can be of the order of $10^3/C$. Notice that, in most relevant physical cases, the number of Lindblad operators $N$ is of the order of the number of system parties~\cite{Eisert}.

All in all, our method is aimed to simulate a different class of master equations with respect the previous approaches, including non-Markovian quantum dynamics, and it is efficient in the range of times where the exponential $e^{M\bar t}$ may be truncated at some low order. A similar result is achieved by the authors of Ref.~\cite{Eisert}, where they simulate a Lindblad equation via Trotter decomposition. They show that the Trotter error is exponentially large in time, but this exponential can be truncated at some low order by choosing the Trotter time step $\Delta t$ sufficiently small. Our method is qualitatively different, and it can be applied also to analogue quantum simulators where suitable entangled gates are available.

Lastly, we note that this method is also appliable to simulate dynamics under a non-Hermitian Hamiltonian $J=H-i\Gamma$, with $H=H^\dag$, $\Gamma=\Gamma^\dag$. This type of generator  emerges as an effective Hamiltonian in the Feshbach partitioning formalism~\cite{Muga04b}, when one looks for the evolution of the density matrix projected onto a subspace. The new Schr\"odinger equation reads
\begin{equation}
\frac{d\rho}{dt}=-i[H,\rho]+\{\Gamma,\rho\},
\end{equation}
This kind of equation is useful in understanding several phenomena, e.g. scattering processes~\cite{Moiseyev98} and dissipative dynamics~\cite{Plenio98}, or in the study of $PT$-symmetric Hamiltonian~\cite{Bender07}.  Our method consists in considering the non-Hermitian part as a perturbative term. As in the case previously discussed, similar bounds can be easily found (see the supplemental material~\cite{suppl}), and this proves that the method is reliable also in this situation.

In conclusion, we have proposed a method to compute expectation values of observables that evolve according to a generalized Lindblad master equation, requiring only the implementation of its unitary part.  Through the quantum computation of $n$-time correlation functions of the Lindblad operators, we are able to reconstruct the corrections of the dissipative terms to the  unitary quantum evolution without reservoir engineering techniques. We have provided a complete recipe that combines quantum resources and specific theoretical developments   to compute these corrections, and error-bounds quantifying the accuracy of the proposal and defining the cases when the proposed method is efficient. Our technique can be also applied, with small changes, to the quantum simulation of  non-Hermitian Hamiltonians. The presented method provides a general strategy to perform quantum simulations of open systems, Markovian or not,  in a variety of quantum platforms.

\bibliographystyle{apsrev}

\section*{Author Contributions}
R.D.C. did the calculations. R.D.C., J.S.P., A.D.C., E.S. and J.C. contributed to the developing of the ideas, obtention of the results and writing the manuscript.

\section*{ACKNOWLEDGMENTS}
The authors thank I\~nigo L. Egusquiza and \'Angel Rivas for stimulating discussions.
The authors  further acknowledge support from the Alexander von Humboldt Foundation;  Spanish MINECO FIS2012-36673-C03-02; UPV/EHU UFI 11/55; UPV/EHU PhD grant; Basque Government IT472-10;  CCQED, PROMISCE, SCALEQIT European projects,  U.S. Department of Energy through the LANL/LDRD Program and a  LANL J. Robert Oppenheimer fellowship (AdC).
 
\newpage

\section*{Supplemental material for  \\``Quantum Simulation of Dissipative Processes without Reservoir Engineering''}

In this Supplementary information, we provide explicit derivations and additional details about the results  in the main text.

\section{Decomposition in Pauli Operators}
In this section, we discuss the decomposition of the Lindblad operators in an unitary basis. In order to implement the protocol of Ref.~\cite{Pedernales} to compute a general multitime correlation function, we need to decompose a general Lindblad operator $L$ and observable $O$ in Pauli-kind orthogonal matrices $\{ Q_k\}_{k=1}^{d^2}$, where $Q_k$ are both Hermitian and unitaries and $d$ is the dimension of the system. If $d=2^l$ for some integer $l$, then a basis of this kind is the one given by the tensor product of Pauli matrices. Otherwise, it is always possible to embed the problem in a larger Hilbert space, whose dimension is the closest power of $2$ larger than $d$. Thus, we can set $\| Q_k\|_\infty=1$ and $\| Q_k\|_2=\sqrt{d}$, where $\|A\|_2\equiv\sqrt{\text{Tr}\,(A^\dag A)}$ and we have redefined $d$ as the embedding Hilbert space dimension. Here, we prove that if $\| L\|_\infty=1$ and $L=\sum_{k=1}^Mq_kQ_k$ with $M\leq d^2$, then (i) $\sum_{k=1}^M |q_k|\leq \sqrt{M}$. This relation will be useful in the proof of Eq.~\eqref{numer1} of the main text. We first show that $\sum_{k=1}^M |q_k|^2\leq1$:
\begin{align}
\sum_{k=1}^M |q_k|^2=\frac{1}{d}\sum_{k=1}^M |q_k|^2\|Q_k\|_2^2=\frac{1}{d}\left\| \sum_{k=1}^Mq_kQ_k\right\|_2^2=\frac{1}{d}\| L\|_2^2\leq \| L\|_\infty^2=1,
\end{align}
where we have used the orthogonality of the matrices $Q_i$, i.e. $\text{Tr}\,(Q_i^{\dag}Q_j)=\text{Tr}\,(Q_i Q_j)=d\delta_{ij}$.
The relation (i) follows simply from the norm inequality for $M$-dimensional vectors $v$: $\| v\|_1\leq \sqrt{M}\|v\|_2$.

\section{Proof of Equation 9}
In this section, we provide a proof of Eq.~\eqref{numer1} of the main text:
\begin{align}
\left|\sum_{[i_1,\dots,i_n]=1}^N\int dV_n\;\langle A_{[i_1,\cdots,i_n]}(\vec{s})\rangle-\frac{(Nt)^n}{n!|\Omega_n|}\sum_{\Omega_n}\tilde A_{\vec{\omega}}(\vec{t})\right|\leq\delta_n
\end{align}
with probability higher than $1-e^{-\beta}$, provided that $|\Omega_n|>\frac{36M_O^2(2+\beta)}{\delta_n^2}\frac{(2\bar{\gamma} M N t)^{2n}}{n!^2}$. Here, $\bar{\gamma}=\max_{i,s\in[0,t]}|\gamma_i(s)|$, $M=\max_i M_i$ where $M_i$ is defined by the Pauli decomposition of the Lindblad operators $L_i=\sum_{k=1}^{M_i}q^i_kQ^i_{k}$, $M_O$ is the Pauli decomposition of the observable $O$ that we will to measure, $[\vec{\omega},\vec{t}]\in\Omega_n\subset\{[\vec{\omega},\vec{t}]\;|\; \vec{\omega}=[i_1,\dots, i_n], i_k\in[1,N], \vec{t}\in V_n \}$ and $[\vec{\omega},\vec{t}]$ are sampled uniformly and independently, $|\Omega_n|$ is the size of $\Omega_n$, and $\tilde A_{\vec{\omega}}(\vec{t})$ corresponds to single-shot measurements of $A_{\vec{\omega}}(\vec{t})$. Notice that $V_n$ is the integration volume corresponding to the $n$-th order term, and $|V_n|=t^n/n!$. 

First, we write $\tilde A_{\vec{\omega}}(\vec{t})=\langle A_{\vec{\omega}}(\vec{t})\rangle+\tilde \epsilon_{[\vec{\omega},\vec{t}]}$, where $\tilde \epsilon_{[\vec{\omega},\vec{t}]}$ is the shot-noise. Note that, due to the previous identity, $\langle \epsilon_{{[\vec{\omega}, \vec{t} ]}} \rangle = 0.$  We have to bound the following quantity
\begin{align}\label{numer2}
&\left|\sum_{[i_1,\dots,i_n]=1}^N\int dV_n\;\langle A_{[i_1,\dots,i_n]}(\vec{s})\rangle-\frac{N^n|V_n|}{|\Omega_n|}\sum_{\Omega_n}\tilde A_{\vec{\omega}}(\vec{t})\right|\leq \nonumber\\
\quad& \leq\left|\sum_{[i_1,\dots,i_n]=1}^N\int dV_n\;\langle A_{[i_1,\dots,i_n]}(\vec{s})\rangle-\frac{N^n|V_n|}{|\Omega_n|}\sum_{\Omega_n}\langle A_{\vec{\omega}}(\vec{t})\rangle\right|+\left|\frac{N^n|V_n|}{|\Omega_n|}\sum_{\Omega_n}\tilde  \epsilon_{[\vec{\omega},\vec{t}]} \right|.
\end{align}
The first term in the right side of Eq.~\eqref{numer2} is basically the error bound in a Montecarlo integration, while the second term is small as the variance of $\epsilon$ is bounded. Indeed, both quantities can be bounded using the Bernstein inequality~\cite{bernstein}:

\newtheorem*{Bernstein}{Theorem}
\begin{Bernstein}[Bernstein Inequality~\cite{bernstein}] Let $X_1,\dots, X_m$ be independent zero-mean random variables. Suppose ${\mathbb{E}[X_i^2]}\leq \sigma_0^2$ and $|X_i|\leq c$. Then for any $\delta>0$,
\begin{align}
\Pr\left[\left|\sum_{i=1}^mX_i\right|>\delta\right]\leq 2\exp\left({\frac{-\delta^2}{4m\sigma_0^2}}\right),
\end{align}
provided that $\delta\leq 2 m\sigma_0^2/c.$
\end{Bernstein}

To compute the first term in the right-hand side of Eq.~\eqref{numer2},  we sample $[\vec{\omega},\vec{t}]$ uniformly and independently to find that $\mathbb{E}\left[\frac{N^n|V_n|}{|\Omega_n|}\langle A_{\vec{\omega}}(\vec{t})\rangle\right]=\frac{1}{|\Omega_n|}\sum_{i_1,\dots,i_n=1}^N\int dV_n\;\langle A_{[i_1,\dots,i_n]}(\vec{s})\rangle$. We define the quantity $X_{[\vec{\omega},\vec{t}]}\equiv\frac{N^n|V_n|}{|\Omega_n|}\langle A_{\vec{\omega}}(\vec{t})\rangle-\frac{1}{|\Omega_n|}\sum_{i_1,\dots,i_n=1}^N\int dV_n\;\langle A_{[i_1,\dots,i_n]}(\vec{s})\rangle $, and look for an  estimate $\left|\sum_{\Omega_n}X_{[\vec{\omega},\vec{t}]}\right|$, where $\mathbb{E}[X_{[\vec{\omega},\vec{t}]}]=0$. We have that 
\begin{align}
\mathbb{E}[X_{[\vec{\omega},\vec{t}]}^2]&=\frac{1}{|\Omega_n|^2}N^{2n}|V_n|^2\mathbb{E}[\langle A_{\vec{\omega}}(\vec{t})\rangle^2]-\frac{1}{|\Omega_n|^2}\left(\sum_{[i_1,\dots,i_n]=1}^N\int dV_n\;\langle A_{[i_1,\dots,i_n]}(\vec{s})\rangle\right)^2\leq\nonumber \\ 
\quad&\leq\frac{N^{n}|V_n|}{|\Omega_n|^2}\sum_{[i_1,\dots,i_n]=1}^N\int dV_n\;\langle A_{[i_1,\dots,i_n]}(\vec{s})\rangle^2\leq\frac{N^{2n}|V_n|^2}{|\Omega_n|^2}\max_{[i_1,\dots,i_n],\vec{s}}\langle A_{[i_1,\dots,i_n]}(\vec{s})\rangle^2,
\end{align}
where we have used the inequality $(\int dV\;f)^2\leq|V|\int dV\;f^2$. Moreover, we have that 
\begin{align}
|X_{[\vec{\omega},\vec{t}]}|=\frac{1}{|\Omega_n|}\left| N^n|V_n|\langle A_{\vec{\omega}}(\vec{t})\rangle-\sum_{[i_1,\dots,i_n]=1}^N\int dV_n\;\langle A_{[i_1,\dots,i_n]}(\vec{s})\rangle\right|\leq\frac{2N^n|V_n|}{|\Omega_n|}\max_{[i_1,\dots,i_n],\vec{s}}|\langle A_{[i_1,\dots,i_n]}(\vec{s})\rangle|,
\end{align}
where we have used the inequality $|\sum_{i=1}^N\int dV\; f|\leq N|V|\max |f|$.
\\
Now, recall that
\begin{align}\label{defA}
 A_{[i_1,\dots,i_n]}(\vec{s})&\equiv  e^{s_n\mathcal{L}_{H_s}^\dag}\mathcal{L}_D^{s_n,i_n\dag}\dots \mathcal{L}_D^{s_2,i_2\dag}e^{(s_1-s_2)\mathcal{L}^\dag_{H_s}} \mathcal{L}_D^{s_1,i_1\dag}e^{(t-s_1)\mathcal{L}^\dag_{H_s}}O,
\end{align}
where $\mathcal{L}_D^{s,i} \xi\equiv\gamma_{i}^{}(s)\left( L_{i}\xi L_{i}^{\dag}-\frac{1}{2}\{ L^{\dag}_{i} L_{i}, \xi\}\right)$, and $\mathcal{L}^\dag\xi\equiv(\mathcal{L}\xi)^\dag$ for a general superoperator $\mathcal{L}$. It follows that  $\max_{[i_1,\dots,i_n],\vec{s}}\langle A_{[i_1,\dots,i_n]}(\vec{s})\rangle^2\leq(2\bar{\gamma})^{2n}\| O\|_\infty^2\prod_{k=1}^n\|L_{i_k}\|_\infty^4= (2\bar{\gamma})^{2n}$, and $\max_{[i_1,\dots,i_n],\vec{s}}|\langle A_{[i_1,\dots,i_n]}(\vec{s})\rangle|\leq (2\bar{\gamma})^n$, where $\bar{\gamma}=\max_{i,s\in[0,t]}|\gamma_i(s)|$ and we have set $\| O\|_\infty=1$ and $\|L_i\|_\infty=1$. Here, we have used the fact that $\langle A_{[i_1,\dots,i_n]}(\vec{s})\rangle$ is real, the inequality $\left|\text{Tr}\,(AB)\right|^2\leq\|A\|_\infty\|B\|_1$, and the result in Eq.~\eqref{Lc} of the next section. 
Now, we can directly use the Bernstein inequality, obtaining
\begin{align}\label{pr1}
\text{Pr}\left[\left|\sum_{\Omega_n}X_{[\vec{\omega},\vec{t}]}\right|>\delta'\right]\leq 2\exp\left(-{\frac{n!^2|\Omega_n|\delta'^2}{4(2\bar{\gamma} Nt)^{2n}}}\right)\equiv p_1
\end{align}
provided that $\delta'\leq (2\bar{\gamma} Nt)^{n}/n!$, and where we have set $|V_n|=t^n/n!$.

Now, we show that the second term in the right hand side of Eq~\eqref{numer2} can be bounded for all $\Omega_n$. From the definition of $\tilde \epsilon_{[\vec{\omega},\vec{t}]}$, we note that 
\begin{align}
\mathbb{E}\left[\frac{N^n|V_n|}{|\Omega_n|}\tilde \epsilon_{[\vec{\omega},\vec{t}]}\right]=\frac{N^n|V_n|}{|\Omega_n|}\sum_{i}\tilde \epsilon^i_{[\vec{\omega},\vec{t}]}p^i_{[\vec{\omega},\vec{t}]}=\frac{N^n|V_n|}{|\Omega_n|}\left(\sum_{i}\tilde A^i_{\vec{\omega}}(\vec{t})p^i_{[\vec{\omega},\vec{t}]}-\langle A_{\vec{\omega}}(\vec{t})\rangle\right)=0,
\end{align} 
where $\tilde \epsilon^{i}_{[\vec{\omega},\vec{t}]}$ ($\tilde A^{i}_{\vec{\omega}}(\vec{t})$) is a particular value that the random variable $\tilde \epsilon_{[\vec{\omega},\vec{t}]}$ ($\tilde A_{\vec{\omega}}(\vec{t})$) can take, and $p^i_{[\vec{\omega},\vec{t}]}$ is the corresponding probability. Notice that the possible values of the random variable $\tilde \epsilon_{[\vec{\omega},\vec{t}]}$ depend on the Pauli decomposition of $A_{\vec{\omega}}(\vec{t})$. In fact, $A_{\vec{\omega}}(\vec{t})$ is a sum of $n$-time correlation functions of the Lindblad operators, and our method consists in decomposing each Lindblad operator in Pauli operators (see section I), and then measuring the real and the imaginary part of the corresponding time-correlation functions. As the final result has to be real, eventually we consider only the real part of $\tilde A_{\vec{\omega}}(\vec{t})$, so that also $\tilde \epsilon_{[\vec{\omega},\vec{t}]}$ can take only real values. In the case $n=2$, one of the terms to be measured is
\begin{align}\label{prod}
L_{\omega_2}^\dag(t_2) L_{\omega_1}^\dag(t_1)O(t)L_{\omega_1}(t_1) L_{\omega_2}(t_2)= 
\sum_{l=1}^{M_O}\sum_{k_1,k_2,k'_1, k'_2=1}^M q_l^Oq_{k_1}^{\omega_1*} q_{k_2}^{\omega_2*}q_{k'_1}^{\omega_1} q_{k'_2}^{\omega_2}\,Q_{k_2}^{\omega_2\dag}(t_2) Q_{k_1}^{\omega_1\dag}(t_1)Q_{l}^{O}(t)Q^{\omega_1}_{k'_1}(t_1) Q^{\omega_2}_{k'_2}(t_2),
\end{align}
where we have used the Pauli decompositions $L_{\omega_i}=\sum_{k_i=1}^{M_{\omega_i}} q_{k_i}^{\omega_i}Q^{\omega_i}_{k_i}$, $O=\sum_{l=1}^{M_O}q_l^OQ_l^O$, and we have defined $M\equiv \max_i M_{\omega_i}$. We will find a bound for the case $n=2$, and the general case will follow straightforwardly. For the term in Eq.~\eqref{prod}, we have that 
\begin{align}\label{bound2}
\sum_{l=1}^{M_O}&\sum_{k_1,k_2,k'_1, k'_2=1}^M |q_l^O|| \Re \;q_{k_1}^{\omega_1*} q_{k_2}^{\omega_2*}q_{k'_1}^{\omega_1} q_{k'_2}^{\omega_2}( \lambda^{\omega_1 \omega_2}_{k_2k_1lk'_1k'_2, r}+i \lambda^{\omega_1\omega_2}_{k_2k_1lk'_1k'_2, im})|\leq\nonumber\\
\quad&\leq 2\sum_{l=1}^{M_O}\sum_{k_1,k_2,k'_1, k'_2=1}^M |q_l^O| | q_{k_1}^{\omega_1*} q_{k_2}^{\omega_2*}q_{k'_1}^{\omega_1} q_{k'_2}^{\omega_2}|\,\| Q_{k_2}^{\omega_2\dag}(t_2) Q_{k_1}^{\omega_1\dag}(t_1)Q_l^O(t)Q^{\omega_1}_{k'_1}(t_1) Q^{\omega_2}_{k'_2}(t_2)\|_\infty\leq \nonumber\\
\quad&\leq2\sum_{l=1}^{M_O}|q_l^O|\sum_{k_1,k_2,k'_1, k'_2=1}^M | q_{k_1}^{\omega_1*} q_{k_2}^{\omega_2*}q_{k'_1}^{\omega_1} q_{k'_2}^{\omega_2}|\leq 2\sqrt{M_O}\,M^2,
\end{align}
where we have defined the real part ($ \lambda^{\omega_1 \omega_2}_{k_2k_1lk'_1k'_2, r}$) and the imaginary part ($\lambda^{\omega_1\omega_2}_{k_2k_1lk'_1k'_2, im}$) of the single-shot measurement of $Q_{k_2}^{\omega_2\dag}(t_2) Q_{k_1}^{\omega_1\dag}(t_1)Q_l^O(t)Q^{\omega_1}_{k'_1}(t_1) Q^{\omega_2}_{k'_2}(t_2)$, and we have used the fact that $\| Q_k^i\|_\infty=1$, $\|Q_l^O\|_\infty=1$, and relation (i) of the previous section. Eq.~\eqref{bound2} is a bound on the outcomes of $L_{\omega_2}^\dag(t_2) L_{\omega_1}^\dag(t_1)O(t)L_{\omega_1}(t_1) L_{\omega_2}(t_2)$.  Notice that the bound in Eq.~\eqref{bound2} neither depends on the particular order of the Pauli operators, nor on the times $s_i$, so it holds for a general term in the sum defining $A_{\vec{\omega}}(\vec{t})$. Thus, we  find that, in the case $n=2$, $\tilde A_{\vec{\omega}}(\vec{t})$ is upper bounded by $|\tilde A_{\vec{\omega}}(\vec{t})|\leq 2\sqrt{M_O}(2\bar{\gamma}M)^2$. In the general case of order $n$, it is easy to show that $|\tilde A_{\vec{\omega}}(\vec{t})|\leq 2\sqrt{M_O}(2\bar{\gamma}M)^n$. It follows that 
\begin{align}
\left|\frac{N^n|V_n|}{|\Omega_n|}\tilde \epsilon_{[\vec{\omega},\vec{t}]}\right|=\frac{N^n|V_n|}{|\Omega_n|}\left|\tilde A_{\vec{\omega}}(\vec{t})-\langle A_{\vec{\omega}}(\vec{t})\rangle\right|\leq \frac{(2\bar\gamma N)^n|V_n|}{|\Omega_n|}(1+2\sqrt{M_O}M^n)\leq \frac{3\sqrt{M_O}(2\bar\gamma M N)^n|V_n|}{|\Omega_n|}.
\end{align}
Regarding the bound on the variance, we have that 
\begin{align}
\mathbb{E}\left[\left(\frac{N^n|V_n|}{|\Omega_n|}\tilde \epsilon_{[\vec{\omega},\vec{t}]}\right)^2\right]&=
\sum_{i}\left(\frac{N^n|V_n|}{|\Omega_n|}\tilde \epsilon^{i}_{[\vec{\omega},\vec{t}]}\right)^2p^i_{[\vec{\omega},\vec{t}]}\leq\frac{N^{2n}|V_n|^2}{|\Omega_n|^2}\sum_i\tilde A^{i\,2}_{\vec{\omega}}(\vec{t})p^i_{[\vec{\omega},\vec{t}]}\leq \nonumber\\
\quad&\leq\frac{N^{2n}|V_n|^2}{|\Omega_n|^2}\max_i \tilde A^{i\,2}_{\vec{\omega}}(\vec{t}) =\frac{N^{2n}|V_n|^2}{|\Omega_n|^2}\left(\max_i |\tilde A^{i}_{\vec{\omega}}(\vec{t})|\right)^2\leq \frac{4M_O(2\bar{\gamma}MN)^{2n}|V_n|^2}{|\Omega_n|^2},
\end{align}
Using Bernstein inequality, we obtain
\begin{align}\label{pr2}
\text{Pr}\left[\left|\frac{N^n|V_n|}{|\Omega_n|}\sum_{\Omega_n}\tilde \epsilon_{[\vec{\omega},\vec{t}]}\right|>\delta''\right]\leq 2\exp\left(- \frac{n!^2|\Omega_n|\delta''^2}{16M_O^2(2\bar{\gamma} MNt)^{2n}}\right)\equiv p_2,
\end{align}
provided that $\delta''\leq \frac{8}{3}\sqrt{M_O}(2\bar{\gamma} MNt)^{n}/n!$, where we have set, as before, $|V_n|=t^n/n!$. Now, choosing $\delta'=\frac{1}{2M^n+1}\delta_n$, $\delta''=\frac{2M^n}{2M^n+1}\delta_n$, $|\Omega_n|>\frac{36M_O^2(2+\beta)}{\delta_n^2}\frac{(2\bar{\gamma} MN t)^{2n}}{n!^2}$, we have that $p_1, p_2\leq \frac{e^{-\beta}}{2}$. Notice that $\delta_n\leq (2\bar{\gamma}Nt)^n/n!$ always holds, so the conditions on $\delta'$, $\delta''$ are satisfied. By using the union bound, we conclude that
\begin{align}
\text{Pr}&\left[ \left|\sum_{[i_1,\dots,i_n]=1}^N\int dV_n\;\langle A_{[i_1,\cdots,i_n]}(\vec{s})\rangle-\frac{(Nt)^n}{n!|\Omega_n|}\sum_{\Omega_n}\tilde A_{\vec{\omega}}(\vec{t})\right|>\delta_n \right]\leq\nonumber\\
&\leq \text{Pr}\left[ \left|\sum_{[i_1,\dots,i_n]=1}^N\int dV_n\;\langle A_{[i_1,\dots,i_n]}(\vec{s})\rangle-\frac{N^n|V_n|}{|\Omega_n|}\sum_{\Omega_n}\langle A_{\vec{\omega}}(\vec{t})\rangle\right|>\frac{1}{1+2M^n}\delta_n \lor \left|\frac{N^n|V_n|}{|\Omega_n|}\sum_{\Omega_n}\tilde  \epsilon_{[\vec{\omega},\vec{t}]} \right| > \frac{2M^n}{1+2M^n}\delta_n\right]\leq \nonumber\\
&\leq p_1+p_2\leq e^{-\beta}.
\end{align}

\section{Proof of Equations 13-14}

In this section, we provide the proof for the bound in Eq.~\eqref{D1zero}, and the general bound in Eq.~\eqref{bounds} of the main text. We note that 
\begin{align}\label{11}
D_1(\rho(t),\tilde{\rho}_0(t))&\leq \frac{1}{2}\int_0^tds\; \|\mathcal{L}_D^s\|_{1\rightarrow1}\|\rho(s)\|_1=\frac{1}{2}\int_0^tds\; \|\mathcal{L}_D^s\|_{1\rightarrow1}
\end{align}
holds, where we have introduced the induced superoperator norm $\| \mathcal{A}\|_{1\rightarrow 1}\equiv\sup_\sigma\frac{\|\mathcal{A}\sigma\|_1}{\|\sigma\|_1}$~\cite{Watrousb}. Moreover, the following bound holds 
\begin{align}\label{Lc}
\|\mathcal{L}_D^t\sigma\|_{1}&=\left\|\sum_{i=1}^N\gamma_i(t)\left(L_i\sigma L_i^\dag-\frac{1}{2}L_i^\dag L_i\sigma-\frac{1}{2}\sigma L_i^\dag L_i\right)\right\|_1\leq \sum_{i=1}^N|\gamma_i(t)|\left(\|L_i \sigma L_i^\dag\|_1+\frac{1}{2}\|L_i^\dag L_i\sigma\|_1+\frac{1}{2}\|\sigma L_i^\dag L_i\|_1\right),\nonumber\\
\quad&\leq 2\sum_{i=1}^N|\gamma_i(t)|\| L_i\|_\infty^2\|\sigma\|_1,
\end{align}
where we have used the triangle inequality and the inequality $\|AB\|_1\leq \left\{\|A\|_\infty\|B\|_1,\|A\|_1\|B\|_\infty\right\}$. Eq.~\eqref{Lc} implies that $\| \mathcal{L}_D^t\|_{1\rightarrow 1}\leq2\sum_{i=1}^N|\gamma_i(t)|\| L_i\|_\infty^2$. Inserting it into Eq.~\eqref{11}, it is found that
\begin{align}
D_1(\rho(t),\tilde{\rho}_0(t))&\leq\sum_{i=1}^N\| L_i\|_\infty^2 \int_0^tds\;|\gamma_i(s)|=\sum_{i=1}^N|\gamma_i(\epsilon_i)|\| L_i\|_\infty^2t,
\end{align}
where we have assumed that $\gamma_i(t)$ are continuous functions in order to use the mean-value theorem ($0\leq\epsilon_i\leq t$). Indeed, $|\gamma_i(\epsilon_i)|=\frac{1}{t}\int_0^tds\;|\gamma_i(s)|$, that can be directly calculated or estimated. 

The bound in Eq.~~\eqref{bounds} has to been proved by induction. Let us assume that Eq.~\eqref{bounds} in the text holds for the order $n-1$. We have that
\begin{align}
D_1(\rho(t),\tilde{\rho}_n(t))&\leq\int_0^tds\;\|\mathcal{L}_D\|_{1\rightarrow1}D_1(\rho(s),\tilde{\rho}_{n-1}(s))\leq \prod_{k=0}^{n-1}\bigg[2\sum_{i_k=1}^N |\gamma_{i_k}(\epsilon_{i_k})|\|L_{i_k}\|_\infty^2\bigg]\sum_{i=1}^N\|L_i\|_\infty^2\frac{1}{n!}\int_0^tds\;|\gamma_i(s)|s^n,
\end{align}
 where we need to evaluate the quantities $\int_0^tds\;|\gamma_i(s)|s^n$. By using the mean-value theorem, we have $\int_0^tds\;\gamma_i(s)s^n=|\gamma_i(\epsilon_i)|\int_0^tds\;s^n$, with $0\leq\epsilon_i\leq t$, and Eq.~\eqref{bounds} follows straightforwardly. In any case, we can evaluate $\int_0^tds\;|\gamma_i(s)|s^n$ by solving directly the integral or we can estimate it by using H\"older's inequalities:
 \begin{equation}
 \int_0^tds\;|\gamma_i(s)|s^n\leq\left\{\sqrt{\int_0^tds\;\gamma_i(s)^2}\;\sqrt{\frac{t^{2n+1}}{2n+1}}\;, \;\max_{0\leq s\leq t}|\gamma_i(s)|\;\frac{t^{n+1}}{n+1}\right\}.
 \end{equation}
 
\section{Error bounds for the expectation value of an observable}
 
 In this section, we find an error bound for the expectation value of a particular observable $O$. As figure of merit, we choose $D_O(\rho_1,\rho_2)\equiv \left|\text{Tr}\,(O(\rho_1-\rho_2))\right|/ (2\| O\|_\infty)$. The quantity $D_O(\rho_1,\rho_2)$ tells us how close the expectation value of $O$ on $\rho_1$ is to the expectation value of $O$ on $\rho_2$, and it is always bounded by the trace distance, i.e. $D_O(\rho_1,\rho_2)\leq D_1(\rho_1,\rho_2)$. Taking the expectation value of $O$ in both sides of Eq.~\eqref{generaldif} of the main text, we find that 
 \begin{align}\label{boundO}
 D_O(\rho(t),\tilde{\rho}_n(t))&=\frac{1}{2\|O\|_\infty}\left|\int_0^tds\;\text{Tr}\,\left(e^{(t-s)\mathcal{L}_H}\mathcal{L}_D^s(\rho(s)-\tilde{\rho}_{n-1}(s))O\right) \right|= \frac{1}{2\|O\|_\infty}\left|\int_0^tds\;\text{Tr}\; \left(\mathcal{L}_D^{s\dag} O(\rho(s)-\tilde{\rho}_{n-1}(s))\right)\right|\leq \nonumber\\
\quad&\leq\frac{1}{\|O\|_\infty}\int_0^tds\;\|\mathcal{L}_D^{s\dag} O\|_\infty D_1(\rho(s),\tilde{\rho}_{n-1}(s))\leq\frac{\|\mathcal{L}_D^{s\dag} O\|_\infty}{\|O\|_\infty}\left(2\bar \gamma N\right)^{n}\frac{t^{n+1}}{2(n+1)!},
\end{align}
where  $\mathcal{L}_D^{s\dag} O=\sum_{i=1}^N\gamma_i(s)\left(L_i^\dag O L_i-\frac{1}{2}\{L_i^\dag L_i ,O\}\right)$. The bound in Eq.~\eqref{boundO} is particularly useful when $L_i$ and $O$ have a tensor product structure. In fact, in this case, the quantity $\|\mathcal{L}_D^{s\dag} O\|_\infty$ can be easily calculated or bounded. For example, consider a $2$-qubit system with $L_1=\sigma^-\otimes \mathbb{I}$, $L_2=\mathbb{I}\otimes \sigma^-$, $\gamma_i(s)=\gamma>0$ and the observable $O=\sigma_z\otimes \mathbb{I}$. Simple algebra leads to $\|\mathcal{L}_D^{s\dag} O\|_\infty=\gamma\| (\mathbb{I}+\sigma_z)\otimes \mathbb{I}\|_\infty=\gamma\|\mathbb{I}+\sigma_z\|_\infty\|\mathbb{I}\|_\infty=2\gamma$, where we have used the identity $\|A\otimes B\|_\infty=\|A\|_\infty\|B\|_\infty$. 

\section{Total number of measurements}

In this section, we provide a magnitude for the scaling of the number of measurements needed to simulate a certain dynamics with a given error $\varepsilon$ and for a time $t$. We have proved that 
\begin{align}
\varepsilon'\equiv D_1(\rho(t),\tilde \rho_n(t))\leq\frac{(2\bar \gamma Nt)^{n+1}}{2(n+1)!},\end{align}
where $\bar \gamma \equiv \max_i |\gamma_i|$.
We want to establish at which order $K$ we have to truncate in order to have an error $\varepsilon'$ in the trace distance. We have that, if $n\geq ex+\log\frac{1}{\tilde\varepsilon}$, with $x\geq 0$ and $\tilde \varepsilon \leq 1$, then $\frac{x^n}{n!}\leq \tilde \varepsilon$. In fact 
\begin{align}
\frac{x^n}{n!}\leq \left(\frac{ex}{n}\right)^n\leq\left(1+\frac{\log\frac{1}{\tilde \varepsilon}}{ex}\right)^{-ex-\log\frac{1}{\tilde \varepsilon}}\leq\left(1+\frac{\log\frac{1}{\tilde \varepsilon}}{ex}\right)^{-ex}\leq e^{-\log \frac{1}{\tilde \varepsilon}}= \tilde \varepsilon,
\end{align}
where we have used the Stirling inequality $n!\geq \sqrt{2\pi n} \,(n/e)^n\geq (n/e)^n$. This implies that, if we truncate at the order $K\geq 2e\bar \gamma Nt+\log\frac{1}{2\varepsilon'}-1=O(2e\bar \gamma Nt+\log\frac{1}{\varepsilon'})$, then we have an error lower than $\varepsilon'$ in the trace distance. The total number of measurements in order to apply the protocol up to an error $\varepsilon'+\sum_{n=0}^K\delta_n$ is bounded by $\sum_{n=0}^K3^n|\Omega_n|$. If we choose $\varepsilon'=c\varepsilon$, $\delta_n=(1-c)\frac{\varepsilon}{(K+1)}$ ($0< c<1$), we have that the total number of measurements to simulate the dynamics at time $t$ up to an error  $\varepsilon$ is bounded by
\begin{align}
\sum_{n=0}^K 3^n|\Omega_n|&=\frac{36M_O^2(2+\beta)(1+K)^2}{(1-c)^2\varepsilon^2}\sum_{n=0}^K\frac{(6\bar \gamma N M t)^{2n}}{n!^2}\leq \nonumber\\
\quad&\leq\frac{36M_O^2(2+\beta)(1+K)^2}{(1-c)^2\varepsilon^2} e^{12\gamma N M t}=O\left(\left(6\bar t+\log\frac{1}{\varepsilon}\right)^2\frac{e^{12M\bar t}}{\varepsilon^2}\right),
\end{align}
where we have defined $\bar t=\bar \gamma N t$.

\section{Bounds for the non-Hermitian Hamiltonian case}

The previous bounds apply as well to the  simulation of a non-Hermitian Hamiltonian $J=H-i\Gamma$, with $H$ and $\Gamma$ Hermitian operators. In this case, the Schr\"odinger  equation reads
\begin{equation}\label{masterc}
\frac{d\rho}{dt}=-i[H,\rho]-\{\Gamma,\rho\}=(\mathcal{L}_H+\mathcal{L}_\Gamma)\rho,
\end{equation}
where $\mathcal{L}_\Gamma$ is defined by $\mathcal{L}_{\Gamma}\,\sigma\equiv-\{\Gamma,\sigma\}$.
Our method consists in considering $\mathcal{L}_\Gamma$ as a perturbative term. To ascertain the reliability of the method, we have to show that bounds similar to those in Eqs.~\eqref{D1zero}-\eqref{bounds} of the main text hold. Indeed, after finding a bound for $\|\rho(t)\|_1$ and $\|\mathcal{L}_{\Gamma}\|_{1\rightarrow1}$, the result follows by induction, as in the previous case.

For a pure state, the Schr\"odinger equation for the projected wavefuntion reads~\cite{Muga04b}
\begin{equation}\label{fesch}
\frac{dP\psi(t)}{dt}=-iP{\bf H}P\psi(t)-\int_0^t\,dsP{\bf H}Qe^{-iQ{\bf H}Qs}Q{\bf H}P\psi(t-s),
\end{equation}
where $P+Q=\mathbb{I}$ and ${\bf H}$ is the Hamiltonian of the total system. One can expand $\psi(t-s)$ in powers of $s$, i.e. $\psi(t-s)=\sum_{n=0}^\infty\frac{(-s)^n}{n!}\psi^{(n)}(t)$, and then truncate the series to a certain order, depending on how fast $e^{-iQ{\bf H}Qs}$ changes. Finally one can find, by iterative substitution, an equation of the kind $d P\psi(t)/dt=JP\psi(t)$, and generalise it to the density matrix case, achieving the equation~\eqref{masterc}, where $\rho$ is the density matrix of the projected system. If the truncation is appropriately done, then we always have $\|\rho(t)\|_1\leq1$ $\forall t\geq0$ by construction. For instance, in the Markovian limit, the integral in Eq.~\eqref{fesch} has a contribution only for $s=0$, and we reach an effective Hamiltonian $J=P{\bf H}P-\frac{i}{2}P{\bf H}Q{\bf H}P\equiv H-i\Gamma$. Here, $\Gamma$ is positive semidefinite, and $\|\rho(t)\|_1$ can only decrease in time.
 
Now, one can easily find that 
\begin{equation}
\|\mathcal{L}_{\Gamma}\sigma\|_1\leq 2\|\Gamma\|_\infty\|\sigma\|_1.
\end{equation}
Hence, $\|\mathcal{L}_{\Gamma}\|_{1\rightarrow1}\leq2\|\Gamma\|_\infty$. 

With these two bounds, it follows that 
\begin{align}
D_1(\rho(t),\tilde{\rho}_0(t))\leq \frac{1}{2}\int_0^tds\; \|\mathcal{L}_\Gamma\|_{1\rightarrow1}\|\rho(s)\|_1\leq\frac{1}{2}\int_0^tds\; \|\mathcal{L}_\Gamma\|_{1\rightarrow1}\leq \|\Gamma\|_\infty t.
\end{align}
One can find bounds for an arbitrary perturbative order by induction, as in the dissipative case.

\end{document}